# CHEMICAL ELEMENT MAPPING BY X-RAY GHOST FLUORESCENCE

Y. Klein[1]*, O. Sefi[1], H. Schwartz[1], and S. Shwartz[1]**.

[1]Physics Department and Institute of Nanotechnology and advanced Materials, Bar Ilan University, Ramat Gan, 52900 Israel.

*yishaykl@gmail.co.il

**sharon.shwartz@biu.ac.il

**Abstract:** *Chemical element mapping is an imaging tool that provides essential information on composite materials and it is crucial for a broad range of fields ranging from fundamental science to numerous applications. Methods that exploit x-ray fluorescence are very advantageous and are widely used, but require focusing of the input beam and raster scanning of the sample. Thus the methods are slow and exhibit limited resolution due to focusing challenges. We demonstrate a new focusing free x-ray fluorescence method based ghost imaging that overcomes those limitations. We combine our procedure with compressed sensing to reduce the measurement time and the exposure to radiation by more than 80%. Since our method does not require focusing, it opens the possibility for improving the resolution and image quality of chemical element maps with tabletop x-ray sources and for extending the applicability of x-ray fluorescence detection to new fields such as medical imaging and homeland security applications.*

X-ray fluorescence (XRF) is a powerful method for the identification and mapping of the chemical compositions of samples with intriguing applications that are exploited in a broad range of fields from fundamental science to industry and cultural heritage. Examples for scientific disciplines where XRF plays a prominent role include materials science, electrochemistry (*1*), biology (*2*), paleontology (*3*), and archeology (*4*). Industrial applications include, for example, metal analyzers for small parts that are produced by the automotive and aerospace industries (*5*). In cultural heritage XRF is very useful in providing information on hidden layers of famous paintings (*6*).

The basic principle of XRF is simple and is based on the x-ray fluorescence process in which x-ray radiation is used to excite core electrons in the sample. When the core electrons are excited or ejected from the inner shells of the atoms, holes are formed in those shells. The electrons can return to their ground state or outer electrons can fill the holes leading to the emission of x-ray radiation at photon energies that correspond to the characteristic atomic lines. The spectrum of the emitted radiation (the fluorescence spectrum) is detected and analyzed, and since each chemical element has unique emission lines, the fluorescence spectrum is used for the characterization of the elemental composition of the sample. The detection is done by energy resolving detectors, which are simple to use, and available components.

While in its simplest form XRF provides no spatial information since the detector collects the radiation from large areas, in recent decades spatially resolved XRF techniques have been developed and their advent opens appealing opportunities in many fields (*1*, *3*, *4*, *6*). However, the main challenge for spatially resolved XRF measurements is that in contrast to transmission measurements the fluorescence is nondirectional, thus the application of pixelated detectors is not trivial. Instead, two-dimensional chemical maps are reconstructed by focusing the impinging beam and raster-scanning the sample. With this procedure, the spatial information is retrieved since at each measurement point only a small portion of the sample is irradiated and the resolution is determined by the spot size of the input beam (*7*). When the spot size is on the order of several microns the method is called micro-XRF. Extensions to three dimensions are also possible by either computed tomography (*8*, *9*) or confocal x-ray microscopy (*10*, *11*), but their implementation is rather challenging.

Despite being very successful and widely used, XRF faces three major challenges that hamper further enhancements of its performances and the extension of its applicability to further disciplines: 1) focusing of x-ray radiation is a challenging task, especially at high photon energies, thus the ability to use small spot sizes in abroad photon energy range is unique to very few synchrotron beamlines and x-ray free electron lasers (*7*). Up to date, the highest resolution achieved using tabletop sources and x-ray capillary lenses is several microns (*12*). However, it is achievable only in a very limited range of photon energies and at the expense of a significant loss of the input flux. 2) In almost all practical implementations of micro-XRF the spatial information is obtained by raster scanning. This is a very slow process since the scan is done over every point of sample. Is clear that the higher the number of required measurements points, the longer the measurement time. For large samples and for three-dimensional imaging the measurement time is several days. 3) For a large group of samples such as biological samples and other materials that are sensitive to x-ray radiation, the currently used levels of dose in micro-XRF are too high, since the radiation causes damage, and their reduction would significantly broaden the applicability and availability of method.

We note that several methods for full field XRF that use photon energy resolving pixelated detectors and that can provide two-dimensional chemical maps in a single frame have been reported (*13*, *14*). However, the spatial resolution is limited to about 10 lines per mm (*13*), which is much worse than the resolution with focusing devices and the quantum efficiency of the detectors drops very quickly at photon energies higher than 20 keV. Moreover, since the fluorescence emission is nondirectional the detector has to be mounted proximally to the sample and cannot provide three-dimensional information without additional lenses (*14*), but this addition introduces severe challenges, and the performances of the system are very limited.

Here we propose and demonstrate a proof of principle experiment for a new, fast, with potentially high-spatial-resolution XRF approach that solves those challenges by using structured illumination and correlation. The main advantages of this approach are that it does not require focusing and that the measurement time can be significantly reduced by using compressed sensing (CS) or artificial intelligence (AI) algorithms.

Our approach is related to the quasi-thermal ghost imaging (GI) approach, which has been investigated extensively in a broad range of wavelengths (*15–32*) from radio waves (*30*) to x-rays (*17–24*), and even with atoms (*33*), neutrons (*34*), and electrons (*35*). GI can be used for the reconstruction of two-dimensional and three-dimensional images (*16*, *24*), and by using CS (*28*) or AI (*29*) the measurement time can be reduced significantly. In the present work, we replace the measurement of the transmission or reflection of the object by the measurement of the x-ray fluorescence, which carries the information on the chemical elements, hence we can use it for chemical mapping.

We note that the GI approach has been used in the visible range for the measurement of the fluorescence (*25–27*). However, long wavelength fluorescence measurements are not element specific and in contrast to our method cannot be used for element mapping. Moreover, the implementation of GI for XRF, where the main alternative is raster scanning, expresses the strengths of the GI approach, which are the abilities to provide spatial information without lenses or mirrors and the natural suitability for compressive measurements, which can be used for the reduction of the measurement time and the dose (*36*).

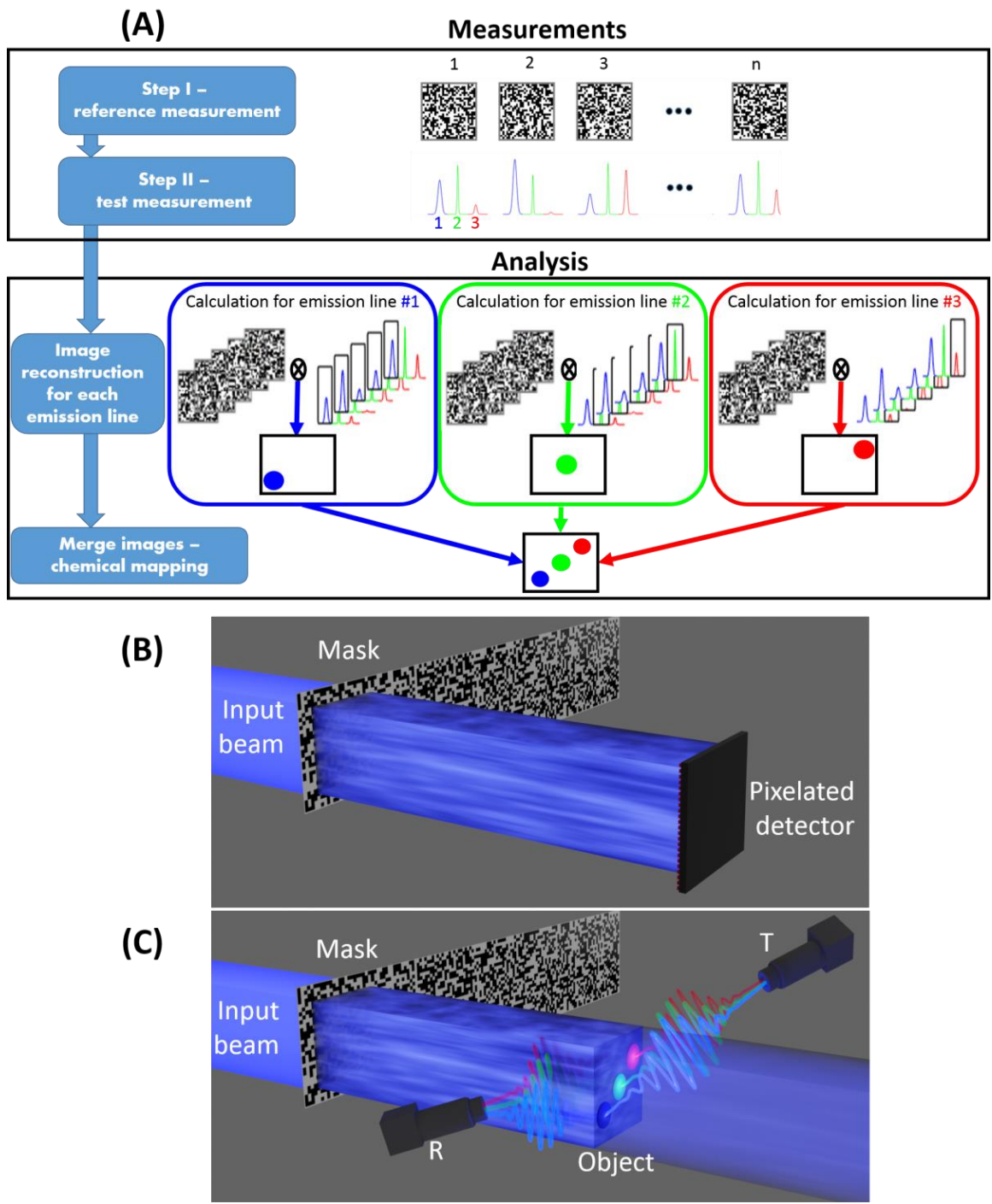

*Figure 1- The image recostruction procedure and schematics of the experimental setup. (A) A flowchart of the reconstruction procedure. In step I, we measure the intensity patterns induced by the mask in the absent of the object. In step II we measure the fluorescence from the object. Schematics of the experimental setup for steps I and II are shown in (B) and (C) respectively.*

Our procedure relies on the two-step approach for the implementation computational GI (*23*). A flow chart that illustrates the procedure is shown in Fig. 1(A). In both steps, the x-ray beam irradiates a mask that induces intensity fluctuations in the beam. The goal of step I is to measure the intensity fluctuations that the mask introduces at the plane of the sample for each of the realizations that we use in step II. As we illustrate in Fig 1(B), this is done in the absence of the object by mounting a pixelated detector at the plane at which we mount object in step II. In step II, which is depicted in Fig 1(C), we remove the pixelated detector, insert the object, and measure the x-ray fluorescence with two photon-energy-resolving silicon drift detectors (SDDs) located at two different positions as is shown in Fig.1(C). We denote the detector located upstream the sample as detector R and the detector downstream the sample as detector T. We then scan the mask at the same positions as in step I and record the fluorescence spectra, which are provided by the SDDs.

After completing the measurements for the entire set of realizations, we obtained sets of data that contain the patterns of the mask (measured in step I) and the corresponding intensities of each of the fluorescence emission lines (measured in step II). To reconstruct each shape of the emitters that emit the fluorescence lines, we exploited the following reconstruction procedure for each chemical element separately. We represented the spatial distribution of each chemical element by a vector **x**. Another vector **T**, which includes n realizations, represents the intensities of the corresponding emission lines measured by the SDD. The mask patterns are represented by the matrix **A** for which every row is a single realization. The vector **T** is equal to the product of the matrix **A** and the vector **x**:

(1) $$\mathbf{Ax} = \mathbf{T}$$

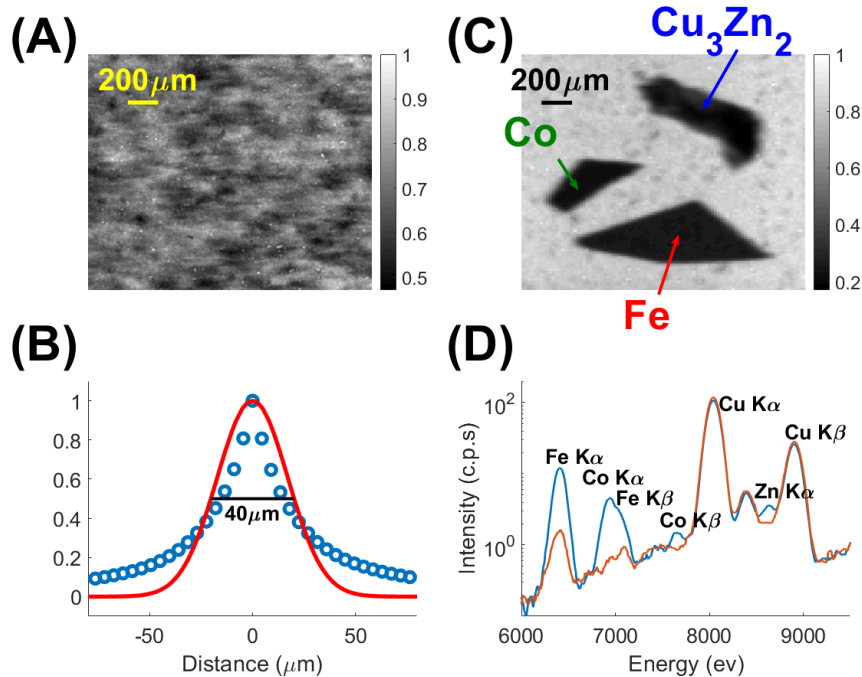

*Figure 2- (A) Example of the reference image (the intesnity fluctuaion induced by the mask). (B) The vertical cross section of the autocorralation of the intensity pattern indcued by the mask averaged over all realizations. The blue dots are the measured data and the red curve is a Gaussian fitting function. (C) Normalized transmission image and (D) Flourescence spectrum of the object, which consist iron, cobalt and brass objects. the red line is the spectrum in the absence of the sample, and the blue line is the spectrum when the sample is present. The emmison lines are indicted near each of the peaks.*

In order to find the vector **x** with a minimal number of realizations, and consequently to reduce the dose and measurement time, we utilized the CS approach, which uses a prior information on the structure of the image. After we reconstructed the image for each of the chemical elements, we overlaid the images to reconstruct the chemical element map.

The source we used for this proof of principle experiment was a rotating copper anode and the mask was a sandpaper with an average feature size of about 45 μm. One example for the speckle pattern, which is detected in step I, is shown in Fig. 2(A). Similar to GI, the spatial resolution of our method is determined by the width of the autocorrelation function of the mask (*37*) that modulates the input x-ray beam. An example for a one-dimensional projection of the autocorrelation function is presented in Fig. 2(B). The autocorrelation function is nearly isotropic, and we found that full width at half maximum (FWHM) of the curve is 40±7 μm in agreement with the average feature size of the mask. The object we imaged contains three small objects made from iron, cobalt, and Brass ($Cu_3Zn_2$). The transmission image and the fluorescence spectrum of the objects are shown in Figs. 2(C) and 2(D), respectively. For the reconstruction of the images, we used the compressed sensing TVAL3 algorithm (*38*).

Our method can provide the chemical map by mounting the detectors at any position around the sample and at any distance as long as they collect the fluorescence as it is emitted from the sample. To demonstrate this ability, we present the images we reconstructed by using our method for the iron and cobalt objects in Figs. 3(A) and 3(B) for detector R and detector T, respectively. The agreement of the chemical element map we reconstructed with the real arrangement and structures of the iron and cobalt objects is excellent and indicates the reliability of our method.

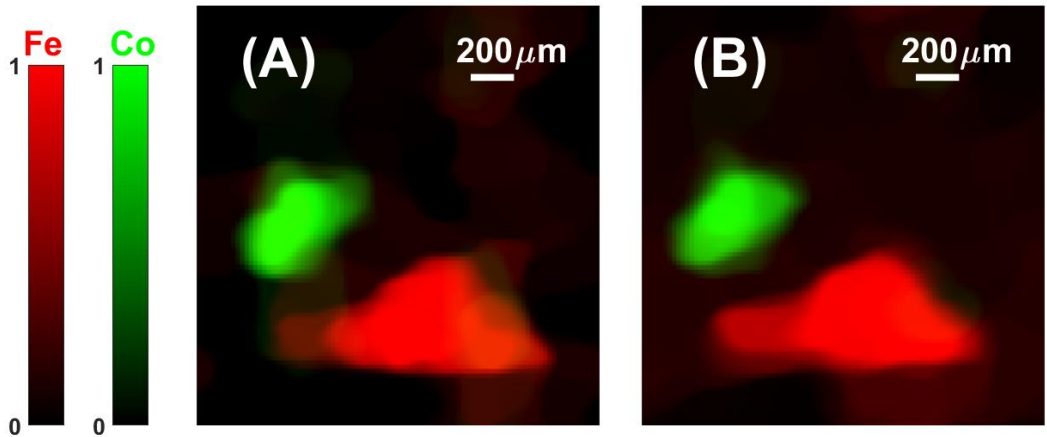

*Figure 3- Reconstructed chemical element maps by x-ray ghost flourescence using (A) detector T and (B) detector R. The red and green areas indicate the areas containing the iron and cobalt elements respectively.*

It is clear that the resolution of our method is much better than the shortest distance between the iron and cobalt objects, which is about 150 microns. This is with agreement with theoretical resolution, which is determined by the width of the autocorrelation function of the mask as discussed above and in contrast to standard micro-XRF for which the spatial resolution is determined by the spot size of the input beam, thus limited by the focusing capabilities. **We therefore are able to overcome one of the greatest challenges of micro-XRF, which is the ability to reconstruct high-resolution chemical maps without focusing.**

The Brass object contains copper and zinc with emission lines for which the photon energy resolution of our detectors is insufficient to distinct them from the characteristic emission lines of our source as can be clearly seen in Fig. 2(D). While for this reason we cannot reconstruct the image of the Brass object, we show very clearly that our method can be used for the elimination of strong background noise and the images of the iron and cobalt objects are very clear despite the strong background (the copper emission lines are stronger than the emission lines of the iron and the cobalt by about a factor of 7).

After demonstrating that our method can provide high-spatial-resolution chemical maps without focusing, we turn to demonstrate that we can utilize CS to reduce number of realizations, hence, to reduce the measurement time and the dose. We plot the reconstructed images that we measured with detector T for various compression ratios (CR) in Fig. 4. The CR is defined by the number of pixels in the map divided by the number of realizations we utilized for the reconstruction (*35*). The maps we described here contains 1010 pixels, which is also the number of sampling points if we were using standard micro-XRF. Consequently, the CR expresses the reduction of the measurement time that our method provides. With our technique we can see a clear image even after 144 realizations, which corresponds to a CR of 7 and identify the objects even with a CR of 20. **The important consequence of this result is that with our method the chemical maps can be reconstructed in much shorter times and with significantly reduced dose compared to standard micro-XRF methods.**

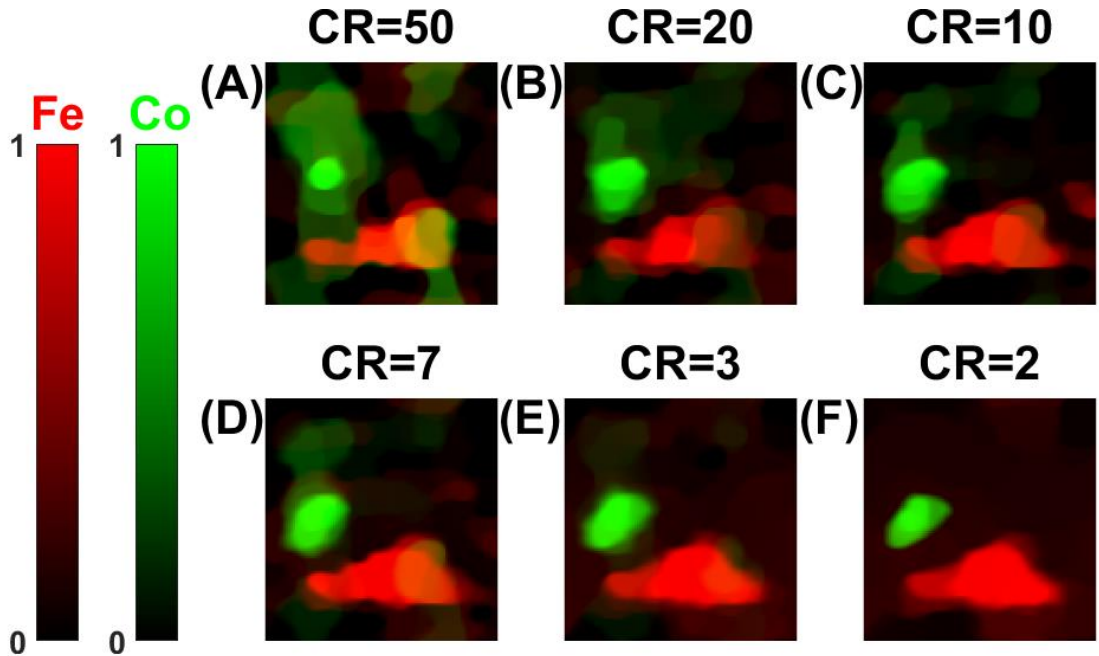

*Figure 4 – Compressive chemical element maps for various compression ratio (CR) values, measured with detector T: (A) CR=50, (B) CR=20, (C) CR=10, (D) CR=7, (E) CR=3, and (F) CR=2.*

Our work opens the possibility for the development of a fast low-dose high-resolution chemical element mapping technique without focusing and without moving the sample. Further generalization of our results will lead to new applications that will extend the capabilities and the impact of XRF to new areas. For example, medical imaging, which is performed at photon energies where lenses are not practical and where the low contrast between various tissues is the main challenge, can benefit from our method. Today, to improve the visibility and quality of images of soft tissues, contrast agents are used since their transmission contrast is higher than the transmission contrast between different types of soft tissues. However, even with the contrast agents the visibility is limited. If instead we can use our method to measure the fluorescence from the same contrast agents, we could increase the quality of the images or alternatively reduce the dose of the measurements, since the fluorescence contrast is significantly higher than the transmission contrast. Another example is for full body scanners used for national security applications; since our method can provide element specific images and can be tuned to be blind to human tissues, it can be used to improve privacy protection of inspected passengers in contrast to other x-ray modalities. Finally, we point out that it is possible to replace the input x-ray beam with an electron beam to excite the inner shell electrons (*39*). In this case spatial resolutions that exceeds the nanometer scale are feasible, and with our method it will be possible to reduce the scanning duration significantly.

**Funding:** Israel Science Foundation (ISF) (201/17); Y.K. gratefully acknowledges the support of the Ministry of Science & Technologies, Israel;

# *Supplementary Materials*

Input beam

The spot size at the plane of the object is about 1.5x1.5 mm$^2$. The spectrum of the input beam consists mainly the copper emission lines, and the Bremsstrahlung x-ray radiation centered around 15 keV. In Fig S1(A) we show the whole input spectrum, from which the red line in Fig. 2(D) is taken.

Autocorrelation function

To estimate the correlation width of the mask, which determines the resolution of our method, we calculated first the correlation width for each of the realizations using the autocorrelation function (*40*):

$$C_i(u,v) = \frac{\sum_{x,y}\left(I_i(x,y)-\bar{I}_i\right)\left(I_i(x-u,y-v)-\bar{I}_i\right)}{\sqrt{\sum_{x,y}\left(I_i(x,y)-\bar{I}_i\right)^2 \sum_{x,y}\left(I_i(x-u,y-v)-\bar{I}_i\right)^2}}$$

For each realization i $I_i(x,y)$ is the intensity at row x and column y and $\bar{I}_i$ is the average intensity. We then averaged the matrix **C** over all the realizations and plot the result in Fig. S1(B). We indicate the cross-vertical section of the autocorrelation function that we plot in Fig. 2(B) of the main text by a green line. While this specific procedure is used to calculate the correlation of the speckle pattern image in the vertical direction, comparable results are obtained for the horizontal direction or for any other direction.

Emission lines

In step II, for the measurement of the single-pixel detector data that correspond to the iron we integrated the intensity over each of the peaks of the Fe $K_\alpha$, $K_\beta$ emission lines, and for the data that correspond to the cobalt we integrated over each the Co $K_\alpha$, $K_\beta$ emission lines. We note that as one can see in Fig. 2(D), there is some overlap between the Co $K_\alpha$ and the Fe $K_\beta$ emission lines, thus we chose only the spectral ranges, which are outside this overlapping region.

Background reduction

Inspecting Fig. S1(A) we can see that the main background sources in our experiment was iron fluorescence from components such as holders and slits. To suppress the strong background from the iron containing materials that are not the object, we measured the emission spectra for all the realizations in the absence of the object and subtracted them from the corresponding spectra that we measured in step II in the presence of the object.

Compress sensing algorithm

To find the vector **x** in Eq. 1 of the main text, which represents the spatial distribution of a chemical element (the corresponding $K_\alpha$, $K_\beta$ emission lines), we used the compress sensing algorithm "total variation minimization by augmented Lagrangian and alternating direction algorithms" (TVAL3) (*37*). The basic idea is to recognize that the gradient of many objects in nature can be represented by a sparse matrix. For each chemical element, the vector **x** is reconstructed by minimizing the augmented Lagrangian:

$$\min_{\mathbf{x}} \sum_{i=1}^{m} \|D_i\mathbf{x}\|_2 \quad s.t \quad \mathbf{Ax=T}, \mathbf{x}>0$$

with respect to the $L_2$ norm. $D_i\mathbf{x}$ is the i$^{\text{th}}$ component of the discrete gradient of the vector **x**.

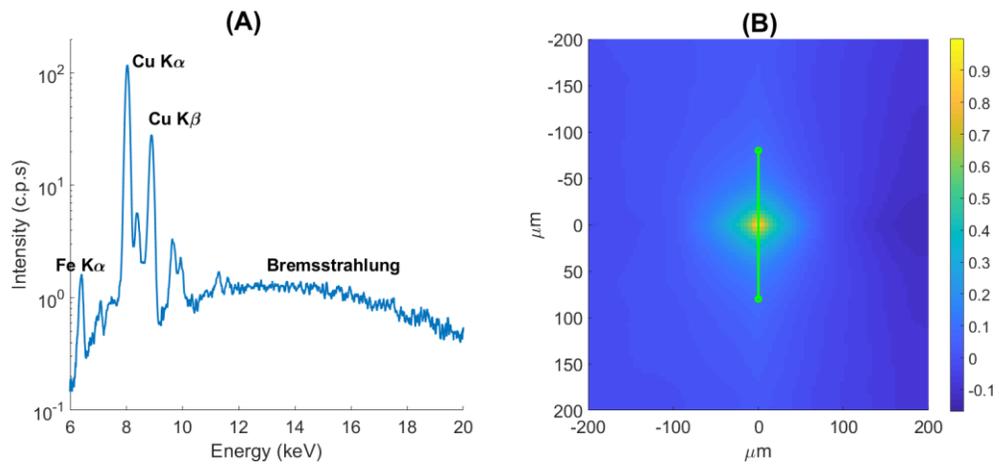

**Fig. S1.**
(A) The spectrum of the input beam. (B) The average autocorrelation matrix (the average is over the realizations).